\documentclass[structabstract]{aa} 
\usepackage{graphicx}
\usepackage{txfonts}
\usepackage{natbib}       

\def\etal{{et\,al.}}

\def\h{$^{\rm h}$}\def\m{$^{\rm m}$}

\def\degs{\ifmmode ^{\circ}\else$^{\circ}$\fi}
\def\fss{\hbox{$.\!\!^{\rm s}$}}        
\def\farcs{\hbox{$.\!\!^{\prime\prime}$}}  

\def\amin{\ifmmode ^{\prime}\else$^{\prime}$\fi}
\def\asec{\ifmmode ^{\prime\prime}\else$^{\prime\prime}$\fi}
\newbox\grsign \setbox\grsign=\hbox{$>$}
\newdimen\grdimen \grdimen=\ht\grsign
\newbox\laxbox \newbox\gaxbox
\setbox\gaxbox=\hbox{\raise.5ex\hbox{$>$}\llap
     {\lower.5ex\hbox{$\sim$}}}\ht1=\grdimen\dp1=0pt
\setbox\laxbox=\hbox{\raise.5ex\hbox{$<$}\llap
     {\lower.5ex\hbox{$\sim$}}}\ht2=\grdimen\dp2=0pt

\begin{document}

   \title{GROND coverage of the main peak of Gamma-Ray Burst 130925A
   \thanks{Partly based on observations collected at the European Organisation 
  for Astronomical Research in the Southern Hemisphere under ID 092.A-0231(B).}}

   \author{J. Greiner\inst{1,2} \and
           H.-F. Yu\inst{1,2} \and
           T. Kr\"uhler\inst{3} \and
           D. D. Frederiks\inst{4} \and
           A. Beloborodov\inst{5} \and
           P. N. Bhat\inst{6} \and
           J. Bolmer\inst{7,1} \and
           H. van Eerten\inst{1}\thanks{Fellow of the Alexander v. Humboldt 
foundation} \and
%
           R. L. Aptekar\inst{4} \and
           J. Elliott\inst{1} \and
           S. V. Golenetskii\inst{4} \and
           J. F. Graham\inst{1} \and
           K. Hurley\inst{8} \and
           D. A. Kann\inst{1,9} \and
           S. Klose\inst{9} \and
           A. Nicuesa Guelbenzu\inst{9} \and
           A. Rau\inst{1} \and
           P. Schady\inst{1} \and
           S. Schmidl\inst{9} \and
           V. Sudilovsky\inst{1} \and
           D. S. Svinkin\inst{4} \and
           M. Tanga\inst{1} \and
           M. V. Ulanov\inst{4} \and
           K. Varela\inst{1} \and
           A. von Kienlin\inst{1} \and
           X.-L. Zhang\inst{1}
          }
   \institute{Max-Planck-Institut f\"ur extraterrestrische Physik,
              Giessenbachstrasse 1, 85748 Garching, Germany
         \and
            Excellence Cluster Universe, Technische Universit\"{a}t M\"{u}nchen,
           Boltzmannstra{\ss}e 2, 85748, Garching, Germany
         \and
            European Southern Observatory, Alonso de C\'{o}rdova 3107, 
            Vitacura, Casilla 19001, Santiago 19, Chile 
         \and
            Ioffe Physical-Technical Institute, Polytekhnicheskaya 26,
            St. Petersburg, 194021 Russia
         \and
            Columbia University, Physics Dept., 538 W 120th Street, New York,
            NY 10027, U.S.A.
         \and
              Center for Space Plasma and Aeronomic Research, Univ. Alabama,
              Huntsville AL 35805, U.S.A.
         \and
              Technische Universit\"at M\"unchen, Physik Dept.,
              James-Franck-Str., 85748 Garching, Germany
          \and
              Space Sciences Laboratory, University of California,
              Berkeley, CA 94720, U.S.A.
         \and
             Th\"uringer Landessternwarte Tautenburg, Sternwarte 5,
             07778 Tautenburg,  Germany 
             }

   \date{Received 22 May 2014; accepted 11 Jul 2014}


  \abstract
   {}
    {Prompt or early optical emission in gamma-ray bursts is notoriously
     difficult to measure, and observations of the dozen cases show a large
     variety of properties. Yet, such early emission promises to help us achieve
     a better understanding of the GRB emission process(es).
    }
   {
    We performed dedicated observations of the ultra-long duration 
    (T90 about 7000 s) GRB 130925A in the 
    optical/near-infrared with the 7-channel 
    ``Gamma-Ray Burst Optical and Near-infrared Detector'' (GROND) at the 
    2.2m MPG/ESO telescope.
   }
    {We detect an optical/NIR flare with an amplitude of nearly 2 mag which
    is delayed with respect to the keV--MeV prompt emission by about 300--400 s.
    The decay time of this flare is shorter than the duration of the flare
    (500 s) or its delay.
    }
    {While we cannot offer a straightforward explanation, we discuss the 
    implications of the flare properties and suggest ways toward 
    understanding it.
    }


   \maketitle
%

\section{Introduction}

\subsection{Early optical emission in GRBs}

\begin{figure}[t]
\vspace{-0.4cm}
\hspace*{-0.5cm}\includegraphics[width=9.6cm]{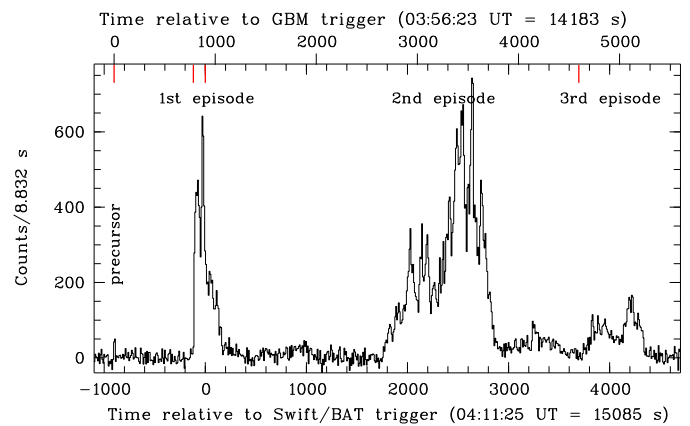}
\vspace{-1.5cm}
\caption{Background-subtracted light curve of the prompt emission of 
GRB 130925A as measured with Konus-Wind in the 26--1480 keV band. 
The red, thick vertical lines at the top indicate, in temporal sequence from
the left, the first GBM trigger, the second GBM trigger, the BAT trigger
and the MAXI trigger.
\label{KWlc}}
\end{figure}

Optical/near-infrared emission from gamma-ray burst sources is very
diverse. Rapid optical observations have shown that the canonical 
power law decay is preceded by a rising part (e.g., \citealt{ryk04},
\citealt{Oates09}, \citealt{mkm10}). On top of this,
some afterglows have shown substantial optical variability, both 
at early times as well as at late times. The early ones can be 
resolved into a component which tracks the prompt gamma-rays
(e.g., GRB 041219A: \citealt{vest05, bbs05}, GRB 050820A: \citealt{vest06}, 
 GRB 080319B: \citealt{rks08})
and an afterglow component which starts during or shortly after the
prompt phase 
(e.g., GRB 990123: \citealt{akerlof99}, GRB 030418: \citealt{ryk04}, 
GRB 060111B: \citealt{kgs06}, GRB 121217A: \citealt{Elliott14}).
The former component has been attributed to internal shocks,
while the latter component has been interpreted as reverse shock emission,
e.g., \cite{sap99, mer99} or 
residual shell collisions at larger radii \citep{liw08}.
Alternative models include a different, optically thin emission
region \citep{Fan2009}, or dust destruction \citep{Cui2013}.
At intermediate times, a large fraction of GRBs has been found to 
exhibit optical flares during the first 1000 s  \citep{Swenson13}. 
At late times, some GRB afterglows (021004, 030329, \citealt{log04})
showed bumps on top of the canonical fading, on timescales of
10$^4$--10$^6$ sec. Originally, these bumps were interpreted as
the interaction of the fireball with moderate density enhancements in the 
ambient medium, with  a density contrast of order 10 \citep{lrc02},
and later by additional energy injection episodes \citep{bgj04, ucg05}.

\subsection{GROND and GRB 130925A}

GROND, a simultaneous 7-channel optical/near-infrared ima\-ger 
\citep{gbc08} moun\-ted at  the 2.2\,m telescope of the 
Max-Planck-Gesellschaft (MPG), operated by MPG 
at the ESO (European Southern Observatory)  La Silla Observatory
(Chile), started operation in May 2007. 
GROND was built as a dedicated GRB and transient 
follow-up instrument
and has observed basically every GRB visible from La Silla 
(weather allowing) since April 2008.
GROND observations of GRBs within the first day are fully automated
(see \citealt{gbc08} for more details). 
The spectral energy distribution (SED) obtained with GROND between
400--2400 nm allows us to not only find high-$z$ candidates 
\citep{gkf09, ksg11}, but also to measure the extinction and the power 
law slope \citep{gkk11} with good accuracy.

GRB 130925A triggered the Gamma-Ray Burst Monitor (GBM, Meegan et al. 2008)
on the \textit{Fermi} satellite first at  03:56 UT on 25 September 2013 
on what seems to be a precursor, and a second time at 04:09 UT 
\citep{Fitzpatrick13}. Subsequently, the Burst Alert Telescope (BAT) on the 
\textit{Swift} satellite \citep{gcg04} triggered at 
$T_0({\rm BAT})$
= 04:11:25 UT = 15085 s (trigger=571830; \citealt{Lien13}). 
Also, INTEGRAL SPI/ACS \citep{Savchenko13} and MAXI/GSC \citep{Suzuki13}
triggered on this gamma-ray burst, and Konus-Wind detected it 
in waiting mode \citep{Golenetskii13}. 
It was also observed by Mars Odyssey and MESSENGER in the interplanetary
network, and these observations strengthen the case for a common origin
for all of the emission episodes \citep{hga13}.
The overall Konus-Wind light 
curve of this particularly long burst, unaffected by Earth occultations, 
is shown in Fig. \ref{KWlc}, with the GBM, BAT and MAXI trigger times
labelled.

The \textit{Swift} satellite slewed immediately upon the BAT trigger
and started taking data with the XRT and UVOT
telescopes at 147 s after the trigger. A bright X-ray source was found 
at RA (2000.0) = 02\h 44\m 42\fss4, 
Decl. (2000.0) = $-$26\degr 09\amin 16\asec\,
with an error radius of 5\farcs1 \citep{Lien13}.
UVOT did not detect any obvious emission, but with GROND a
very red source was detected \citep{Sudilovsky13a}. Subsequent 
spectroscopy with UVES \citep{Vreeswijk13} and 
X-Shooter \citep{Sudilovsky13b} revealed multiple emission lines,
suggestive of a host galaxy redshift of z=0.347.
Using standard cosmology 
($H_{\rm o}$=70 km/s/Mpc, $\Omega_{\rm M}$=0.27,
$\Omega_{\rm \Lambda}$=0.73), this implies a luminosity distance of
0.57$\times$10$^{28}$ cm.

\begin{figure}[ht]
\hspace{-0.9cm}\includegraphics[width=10.3cm]{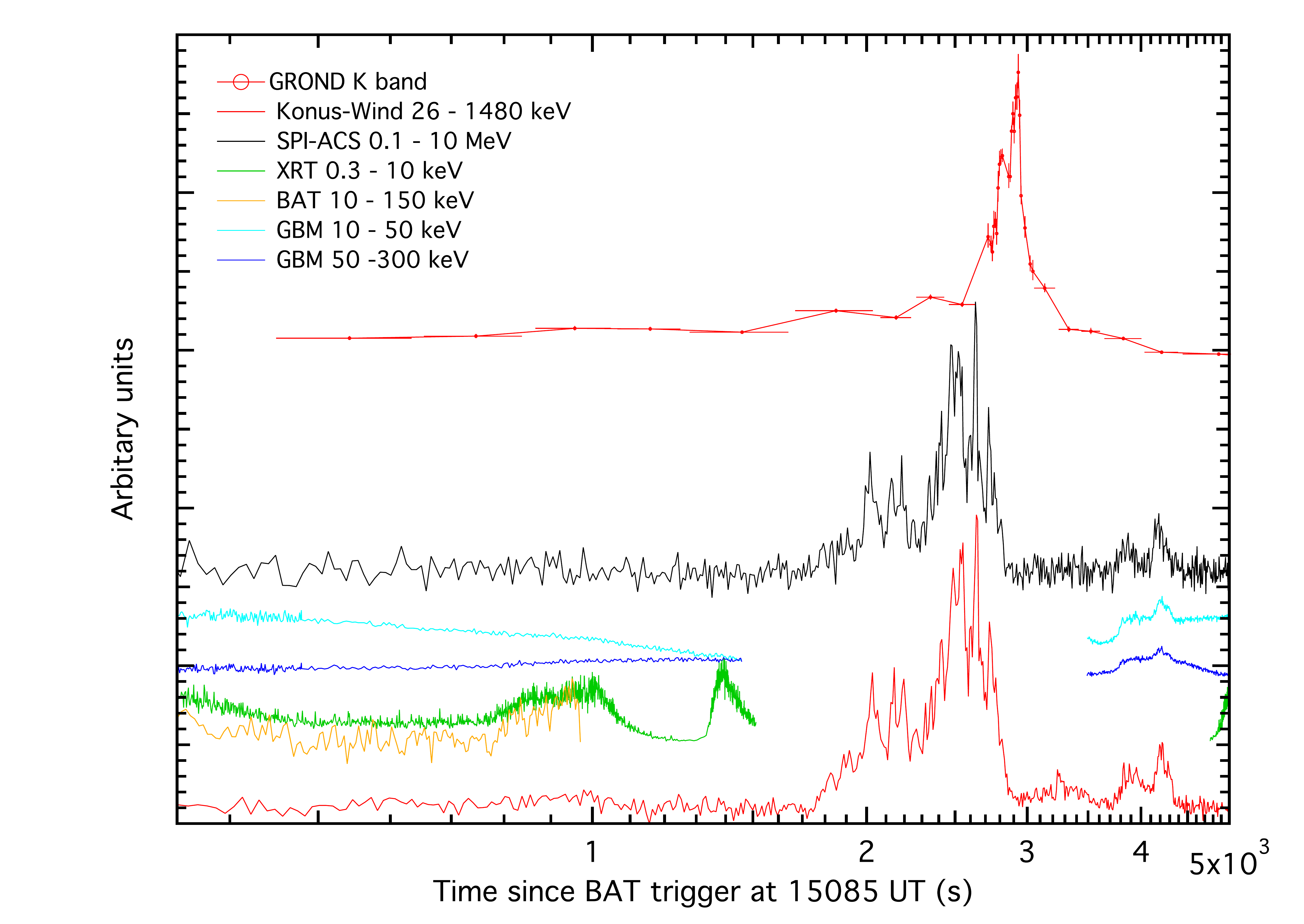}
\caption{Light curve of GRB 130925A as observed with various 
instruments (see labels) at different times. The gap at 
$\approx$ 1.6--3.5 ks for GBM and \textit{Swift}/XRT is
due to Earth blockage.
The top line shows the $K_s$ light curve
as measured with GROND at the same time axis as the gamma-ray emission.
\label{lc}}
\end{figure}

The initial  \textit{Swift}/XRT observations show extremely rapid and 
dramatic flaring, which prompted speculations that this event
could be a tidal disruption event (TDE; \citealt{Burrows13}), similar
to the highly variable soft X-ray light curve observed from 
Swift J1644+57 (GRB 110328A; \citealt{Burrows11, Levan11}).
Late-time HST observations show that the transient emission is
slightly offset from the nucleus of the galaxy by about 0\farcs12 
($\approx$ 600 pc in projection), thus arguing against a TDE
unless the galaxy contains more than one super-massive black hole
\citep{Tanvir13}. Also, the late-time ($>10^5$ s) X-ray behaviour
is markedly different from the spiky light curve of Swift J1644+57
and resembles that of a GRB X-ray afterglow,
so we adopt the GRB classification in this paper. It remains true,
though, that GRB 130925A is among the longest-duration GRBs measured
so far, prompting concerns about the 
maximum possible duration of
the central engine in the standard GRB paradigm \citep{Zhang14}.

Here, we describe observations of the second emission peak, motivated
by the GROND data during this time interval.
Given that both the \textit{Swift} and the \textit{Fermi} satellites
were Earth-blocked during most of the GROND observations, we concentrate
here on the $\gamma$-ray data of the Konus-Wind and INTEGRAL SPI/ACS
instruments, which provide simultaneous coverage. We have
checked RHESSI data, but found it to only exhibit background fluctuations
for this second emission period (D. Smith, priv. comm.).
In the following, our multi-wavelength observations of GRB 130925A are
presented in 
$\S$2, and an explanation of the data is proposed in $\S$3.
Throughout this paper, we use the definition 
$F_{\nu} \propto t^{-\alpha} \nu^{-\beta}$ where $\alpha$ is the 
temporal decay index, and $\beta$ is the spectral slope.

\section{Observations}

\subsection{Konus-Wind data}

The Konus-Wind instrument (KW, \citealt{Aptekar1995}) is
a $\gamma$-ray spectrometer consisting of two identical
detectors, S1 and S2, which observe the Southern and
Northern ecliptic hemispheres, respectively. GRB 130925A was
observed as a count rate increase in the S1 detector.
Thanks to the remote KW orbit around the Lagrangian point L1,
with stable background and lack of Earth or planet occultations,
the instrument was able to measure the burst's prompt emission for 
more than 5000 s.
A more detailed description of the KW observation of GRB 130925A
can be found in \cite{Evans14}.

Since Konus-Wind did not trigger on any of the various peaks, the data
are available only in 'waiting mode'. In this regime, count rates 
with a coarse time resolution of 2.944 s are recorded
in three energy bands: G1 (26--99~keV), G2 (99--394~keV), and 
G3 (394--1480~keV).
The light curve of the event in the combined G1+G2+G3 energy bands
(Fig. \ref{KWlc}) starts with the weak precursor at 
$\sim T_0({\rm BAT})-900$ s
and thereafter shows several multi-peaked pulses, separated by long 
periods of low-level emission.

\begin{table*}[th]
   \caption{Secondary standards (all in AB magnitudes) used for the GROND data.
    For the NIR, different standards had to be used due to the brightness
    of the GRB afterglow.}
   \vspace{-0.2cm}
      \begin{tabular}{cccccccc}
      \hline
      \noalign{\smallskip}
    Filter &  Star I       & Star II       & Star III      & Star IV       & Star V        & Star VI       & Star VII      \\
           & 02:44:44.69   & 02:44:35.90   & 02:44:36.52   & 02:44:34.38   & 02:44:35.89   & 02:44:49.04   & 02:44:42.80   \\
           & $-$26:07:57.0 & $-$26:09:07.3 & $-$26:08:09.3 & $-$26:07:27.2 & $-$26:07:26.9 & $-$26:09:28.8 & $-$26:07:15.4 \\
     \noalign{\smallskip}
     \hline
     \noalign{\smallskip}
  $g'$ & 17.48$\pm$0.04 & 17.65$\pm$0.04 & 21.95$\pm$0.07 & 18.38$\pm$0.04 & 21.05$\pm$0.06 & 20.82$\pm$0.05 & 22.76$\pm$0.08 \\
  $r'$ & 17.06$\pm$0.04 & 17.24$\pm$0.04 & 20.33$\pm$0.05 & 16.92$\pm$0.04 & 19.45$\pm$0.05 & 19.29$\pm$0.04 & 21.30$\pm$0.07 \\
  $i'$ & 16.90$\pm$0.04 & 17.11$\pm$0.04 & 18.71$\pm$0.04 & 16.02$\pm$0.04 & 17.87$\pm$0.05 & 18.20$\pm$0.05 & 19.84$\pm$0.06 \\
  $z'$ & 16.85$\pm$0.04 & 17.04$\pm$0.04 & 18.03$\pm$0.04 & 15.60$\pm$0.04 & 17.12$\pm$0.04 & 17.72$\pm$0.05 & 19.19$\pm$0.06 \\
     \noalign{\smallskip}
     \hline
     \hline
     \noalign{\smallskip}
    Filter &  Star 1 (=I)  & Star 2        & Star 3        & Star 4 (=IV)  & Star 5 (=V)   & Star 6        & Star 7        \\
           & 02:44:44.69   & 02:44:40.97   & 02:44:51.60   & 02:44:34.38   & 02:44:35.89   & 02:44:42.60   & 02:44:51.73   \\
           & $-$26:07:57.0 & $-$26:11:31.8 & $-$26:08:53.9 & $-$26:07:27.2 & $-$26:07:26.9 & $-$26:10:08.4 & $-$26:08:24.0 \\
     \noalign{\smallskip}
     \hline
     \noalign{\smallskip}
   $J$  & 17.03$\pm$0.05 & 16.34$\pm$0.05 & 18.71$\pm$0.07 & 15.44$\pm$0.05 & 16.71$\pm$0.05 & 18.89$\pm$0.07 & 18.22$\pm$0.06 \\
   $H$  & 17.13$\pm$0.06 & 16.13$\pm$0.05 & 18.51$\pm$0.08 & 15.26$\pm$0.05 & 16.50$\pm$0.05 & 18.56$\pm$0.08 & 18.02$\pm$0.07 \\
   $K_{\rm s}$  & 17.50$\pm$0.08 & 16.36$\pm$0.07 & 18.31$\pm$0.09 & 15.49$\pm$0.07 & 16.70$\pm$0.08 & 18.49$\pm$0.10 & 17.90$\pm$0.08 \\
     \noalign{\smallskip}
      \hline
   \end{tabular}
   \label{compstar}
\end{table*}

The time history recorded in the three Konus-Wind energy bands can be 
considered as
a continuous three-channel spectrum covering the 26--1480 keV energy range.
After subtracting the background emission as estimated by a polynomial fit
from before the precursor to well after the third emission period,
we measure the following fluences (all in the standard KW 20--10000 keV band) 
for the different emission episodes based on cut-off powerlaw fits to the 
spectra:
1st major episode (14961--15264 s UT) : 9.5$\pm$0.7$\times$10$^{-5}$ erg cm$^{-2}$;
2nd major episode (16766--17964 s UT) : 3.9$\pm$0.1$\times$10$^{-4}$ erg cm$^{-2}$;
3rd major episode (18768--19463 s UT) : 5.6$\pm$0.9$\times$10$^{-5}$ erg cm$^{-2}$.
The total fluence, which accounts also for the weaker inter-pulse emission,
is 6.2$\pm$0.3$\times$10$^{-4}$ erg cm$^{-2}$.
At $z=0.347$ and standard cosmology 
($H_{\rm o}$=70 km/s/Mpc, $\Omega_{\rm M}$=0.27,
$\Omega_{\rm \Lambda}$=0.73), this implies an isotropic equivalent energy 
release of $E_{\rm iso}$ = 1.9$\pm$0.1 $\times$ 10$^{53}$ erg.
This is one of the most energetic low-redshift GRBs: to our knowledge
it is \#2 after GRB 130427A \citep{Maselli14}.
As it will be important in the later discussion, we also provide the 
individual contributions of the main emission peaks as follows: 
2.9$\pm$0.2$\times$10$^{52}$ erg for 
the first
and 1.2$\pm$0.04$\times$10$^{53}$ erg for 
the second major emission interval.

\subsection{INTEGRAL SPI/ACS data}

The SPI-ACS data were taken from the Integral Data Centre. 
The original 50 ms, 0.1--10 MeV light curve was rebinned into 8 s temporal
bins as shown in Fig. \ref{lc}, but analyzed at their finest time
resolution for the minimum variability timescale.

\subsection{GROND and VLT observations}

\begin{figure}[th]
\vspace{-0.3cm}
\includegraphics[width=9.2cm]{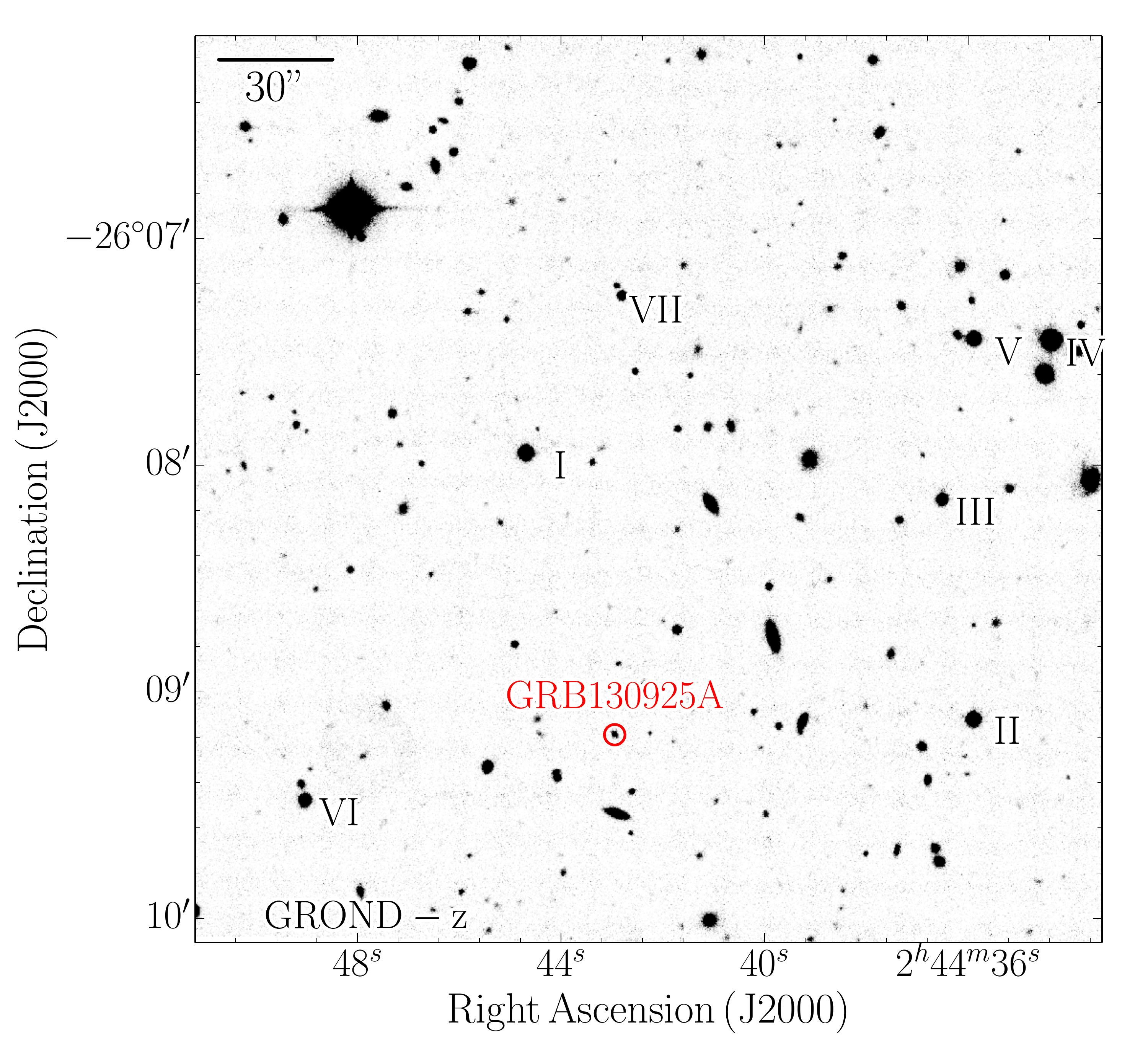}
\vspace{-0.3cm}
\caption{GROND $z'$-band finding chart of GRB 130925A,
including the photometric comparison stars (roman letters) for $g'r'i'z'$
and partly for $JHK_{\rm s}$ as detailed in Tab. \ref{compstar}
(some comparison stars for $JHK_{\rm s}$ 
are outside the field as shown here).
North is up, and East to the left.
\label{grondfc}}
\end{figure}

GROND exposures automatically started 350 s after the 
arrival of the \textit{Swift}/BAT trigger. 
Simultaneous imaging in $g'r'i'z'JHK_{\rm s}$ continued for 
nearly 6 hrs, with the last exposures done only in the NIR 
channels due to morning twilight (after 09:36 UT).
Further observations were done in the mornings of Sep 26, 27
and 29, with one late-time deep host imaging on Oct. 6.
GROND data have been reduced in the standard manner \citep{kkg08}
using pyraf/IRAF \citep{Tody1993, kkg08b}.
The optical/NIR imaging was calibrated against the primary 
SDSS\footnote{http://www.sdss.org} 
standard star network, or catalogued magnitudes of field stars from the 
SDSS in the case of $g^\prime r^\prime i^\prime z^\prime$ observations,
or the 2MASS catalog for $JHK_{\rm s}$ imaging. This results in typical 
absolute accuracies  of $\pm$0.03~mag in $g^\prime r^\prime i^\prime 
z^\prime$ and $\pm$0.05~mag in $JHK_{\rm s}$. 
Comparison stars are given in Tab. \ref{compstar} and labeled in 
the finding chart of GRB 130925A (Fig. \ref{grondfc}).

\begin{figure*}[t]
\includegraphics[width=13.70cm]{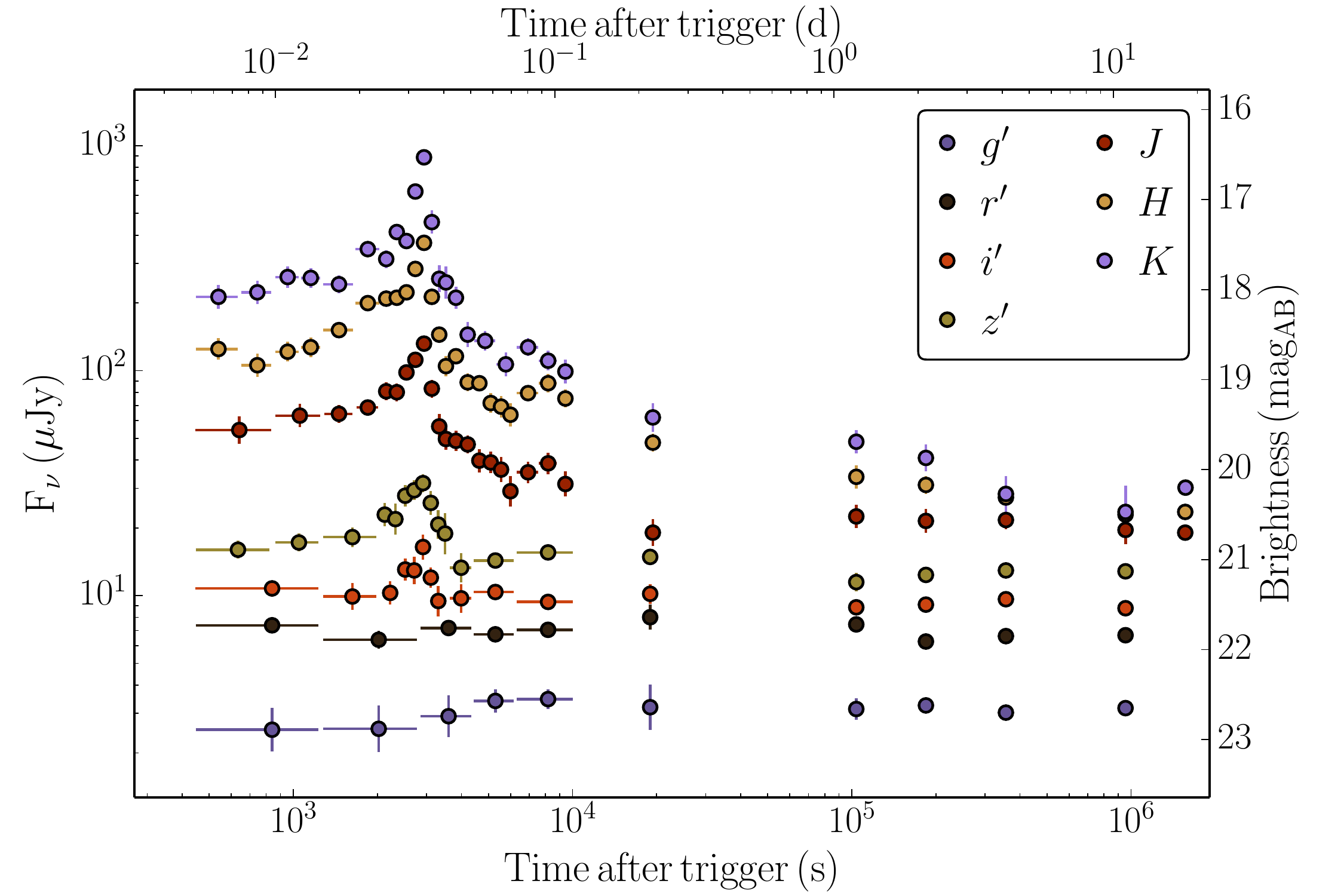}
\caption{Light curve of the afterglow of GRB 130925A in the 
7 GROND filters. The times are relative to the \textit{Swift}/BAT
trigger time. The very last epoch measurements are from HAWK-I
observations and represent our best estimate of the host magnitudes.
\label{grondlc}}
\end{figure*}

We find a variable optical/NIR object at position
RA(2000.0) = 02:44:42.96,
Decl.(2000.0)= --26:09:11.2
($\pm$0\farcs3) which is fully consistent with the \textit{Swift}/XRT 1\farcs4 
error circle (\textit{Swift}/XRT repository).
We therefore identify this object as the optical/NIR counterpart
of GRB 130925A; its light curve in all 7 GROND filters is 
shown in Fig. \ref{grondlc}. 

In order to more accurately estimate the contribution of the
constant host flux to the early- and late-time light curve, we 
observed the field of GRB 130925A with the ESO/VLT UT4 equipped with
HAWK-I. Observations started on October 13, 2013, 06:21 UT, which is
18.1 days after the GRB trigger. They were performed under clear sky
conditions and consist of sets of dithered exposures with a total
integration time of 30, 12, 10 and 16 minutes in the $Y$, $J$, $H$ and
$K_{\rm{s}}$ filter respectively. Data reduction and photometric
calibration was performed in a similar fashion as for the GROND data.
The measurements are shown as the temporally last data points in
Fig. \ref{grondlc}.

When comparing to the GROND light curve,
the most intriguing feature is the long and clearly visible 
(Fig. \ref{lc}) delay of the optical/NIR emission peak relative
to the $\gamma$-ray emission (second episode). 
This delayed optical/NIR emission will be further discussed
in section 3.3. below.

\subsection{Swift data}

The \textit{Swift}/BAT instrument triggered on the first emission episode
(not on the precursor which triggered GBM), and follow-up observations
with the XRT and UVOT instruments started about 150 s after the BAT trigger
\citep{Lien13}. While a 
thorough analysis of all the \textit{Swift} data
has appeared already \citep{Evans14}, we use some \textit{Swift}/XRT 
observations
to compare with the GROND data.
The corresponding data have been extracted from the 
\textit{Swift}/XRT repository \citep{ebp09}.

\section{Results}

\subsection{Minimum variability time scale of the prompt emission}

The variability in the $\gamma$-ray light curve of GRBs is, in most models,
related to a physical origin in the central engine, likely convolved with
the bulk Lorentz factor of the outflow. We have estimated the minimum
variability time scale (MVT) in GRB 130925A with the new method recently
developed for high time-resolution GBM data \citep{Bhat13}. The method
is based on the ratio of variances of the GRB emission and the background
as a function of the bin width.
For the precursor and the first and the third peak as seen by 
\textit{Fermi}/GBM,
we derive the following MVT: 0.39$\pm$0.02 s, 0.92$\pm$0.01 s, 
6.7$\pm$2.2 s, respectively.
For the INTEGRAL/ACS, we obtain 1.10$\pm$0.19 s, 4.82$\pm$1.86 s and 
14.2$\pm$5.2 s for the first, second and third peak, respectively. 
This shows the typical behaviour of increasing MVT with time. 
Given that the INTEGRAL/ACS and \textit{Fermi}/GBM data show good consistency
for the first and third peaks, we can be confident in the 
INTEGRAL-derived MVT value for the second peak, where we have no GBM coverage.
\cite{Bhat13} has estimated the mean
MVT for short- and long-duration GRBs as 0.024 s and 0.25 s, respectively
(see his Fig. 3). Thus, the MVT of about 1 s for the first and about 5 s 
for the second
peak of GRB 130925A falls in the longest 5\%/1\% percentile of the 
distribution of long-duration GRBs.

If the MVT relates to the bulk Lorentz factor, then GRB 130925A
might feature one of the lowest Lorentz factors encountered so far.
Applying the formalism of \cite{lis01}, and carrying forward the error 
of the spectral slope to estimate the flux at 1 MeV, we obtain
minimum Lorentz factors of 37, 20 and 31 for the precursor, first and
third peak as seen with \textit{Fermi}/GBM. We note, though, that using e.g., the
$\Gamma_o$-E$_{\gamma, iso}$ relation of \cite{lyz10}, we obtain values
up to a factor 10 higher.

\subsection{A normal optical/NIR/X-ray afterglow?}

\begin{figure}[th]
\vspace*{-0.4cm}
\includegraphics[width=9.5cm]{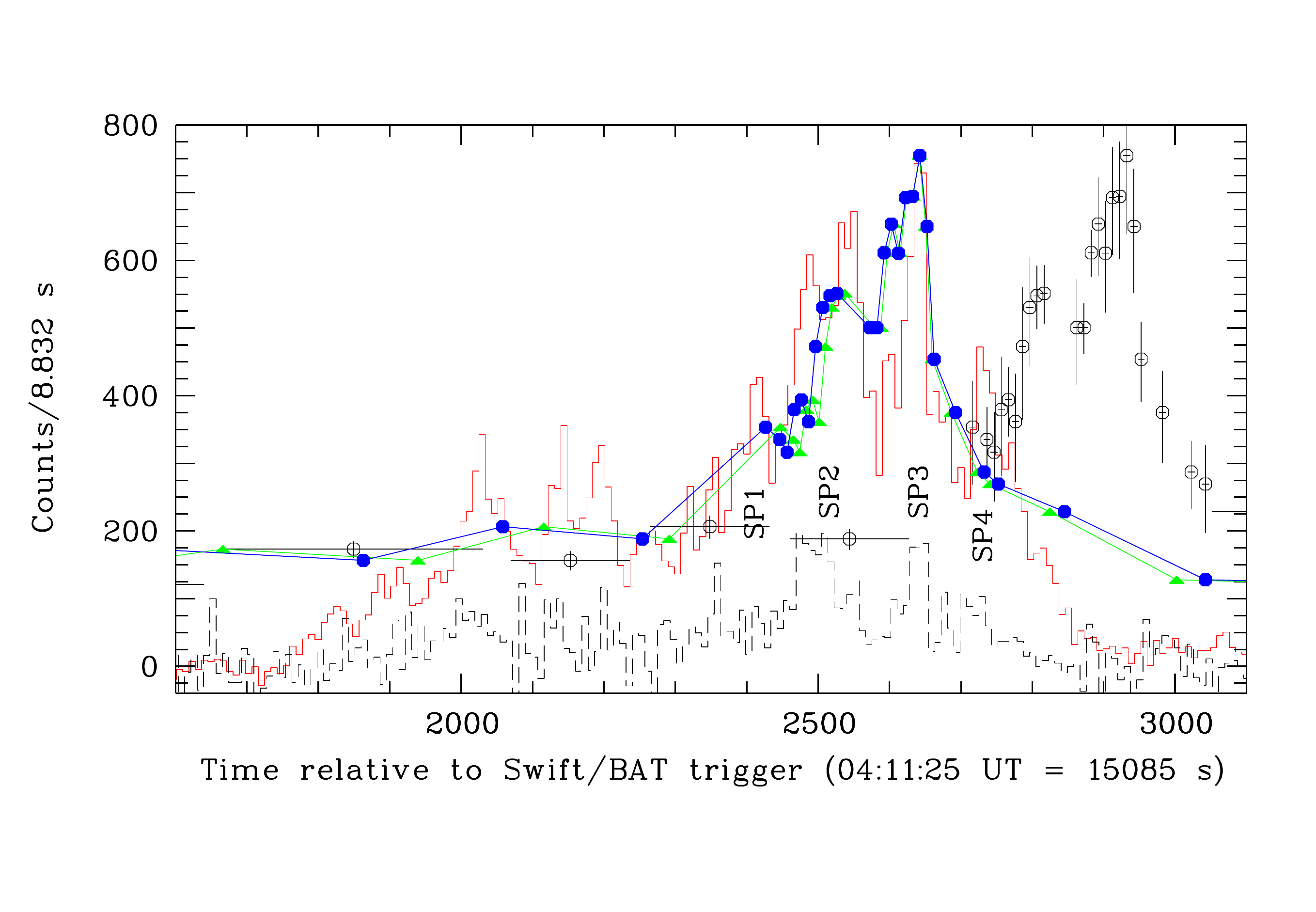}
\vspace*{-1.1cm}
\caption{Expanded view of the delay of the GROND $K_s$-band light curve
(open black symbols with error bars) with respect to the KW light curve
(red line). In blue and green we show the shifted (-290 s) and
stretched (time axis divided by 1.11) GROND $K_s$-band light curve,
respectively. The black dashed curve at the bottom is the KW highest
energy channel G3 (at a factor 3 stretched intensity scaling). 
\label{delay}}
\end{figure}

In an attempt to derive a combined Swift/XRT and GROND SED,
we have selected time intervals for both instruments which were not
affected by flares. Due to the strong flaring activity, Earth blockages
of Swift and observability periods of GROND, we are forced to select
non-simultaneous, not even overlapping time intervals. Using 36 ks
of \textit{Swift}/XRT data from 100--370 ks after the trigger (to 
avoid pile-up in the PC-mode data, but still before the light curve break) 
and GROND data from 3.5--6 ks, we perform a simultaneous 
spectral fit after re-adjusting
the relative normalisation according to the common $t^{-0.9}$ intensity decay.
Since the X-ray spectral slope is substantially steeper than the 
extinction-corrected optical/NIR slope, we employ a broken power law.
In contrast to the typical GRB afterglow SEDs \citep[e.g.,][]{gkk11} a fit
with a fixed slope difference of 0.5 between the optical/NIR and X-ray 
part is rejected at high confidence.
Fixing the redshift and the galactic foreground absorption of 
$N_H^{Gal} = 1.66 \times 10^{20}$ cm$^{-2}$ \citep{kalberla2005}
we obtain the following
best-fit parameters: 
$N_H^{Host} = (1.5\pm0.1) \times 10^{22}$ cm$^{-2}$,
$\beta_{opt/NIR} = 0.32\pm0.03$, $\beta_{X} = 1.6\pm0.1$,
$E_{break} = 1.68\pm0.09$ keV, and
$E(B-V) = 2.24\pm0.26$ mag for SMC-like dust,
at a reduced $\chi^2$=0.87 for 84 degrees of freedom.
While formally an acceptable fit, this is incompatible on physical grounds
with any previous afterglow modelling, and
supports earlier conclusions that suggested a different origin of the 
X-ray emission,
i.e., as dust-scattered prompt flux \citep{Evans14} 
or thermal emission from a hot cocoon \citep{Piro14}.

\subsection{The delayed optical/NIR peak wrt. gamma-rays}

\subsubsection{Temporal properties}

In the GROND data we see a very sudden increase in brightness 
at around 2.5 ks after the BAT trigger, followed 
by a rapid decline
after the peak at 3 ks, followed by a more gentle decline
beyond about 4 ks. There is a pronounced dip around 6 ks
with another emission peak at 8 ks.
Fitting a two-component model with a triangular-shaped flare
on top of a gentle, smoothly-broken powerlaw (as used above) 
for the times before and well after the flare we derive 
a rise slope of $t^{+2.0\pm0.3}$ and
a decay slope of $t^{-6\pm1}$, assuming a $T_o$ at the beginning of the
second gamma-ray pulse.

As already mentioned, the most intriguing feature of GRB 130925A
is the long and clearly visible 
(Fig. \ref{lc}) delay of the optical/NIR emission peak relative
to the $\gamma$-ray emission (second episode). Such lags have
been seen in previous GRBs, though the present delay is surprisingly
long: we measure a delay of the optical/NIR peak 
of 3040$\pm$30 s relative to the maximum of the first gamma-ray episode, 
and 405$\pm$30 s relative to the maximum of the second gamma-ray episode. 

\begin{figure}[ht]
\includegraphics[width=9.1cm]{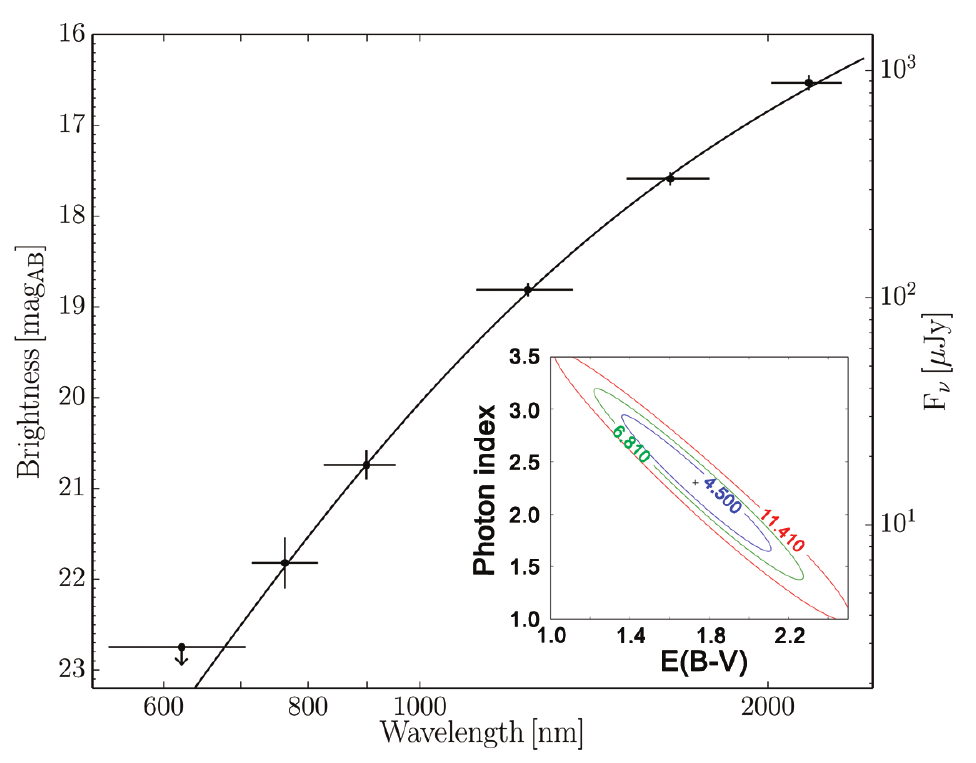}
\includegraphics[width=8.2cm]{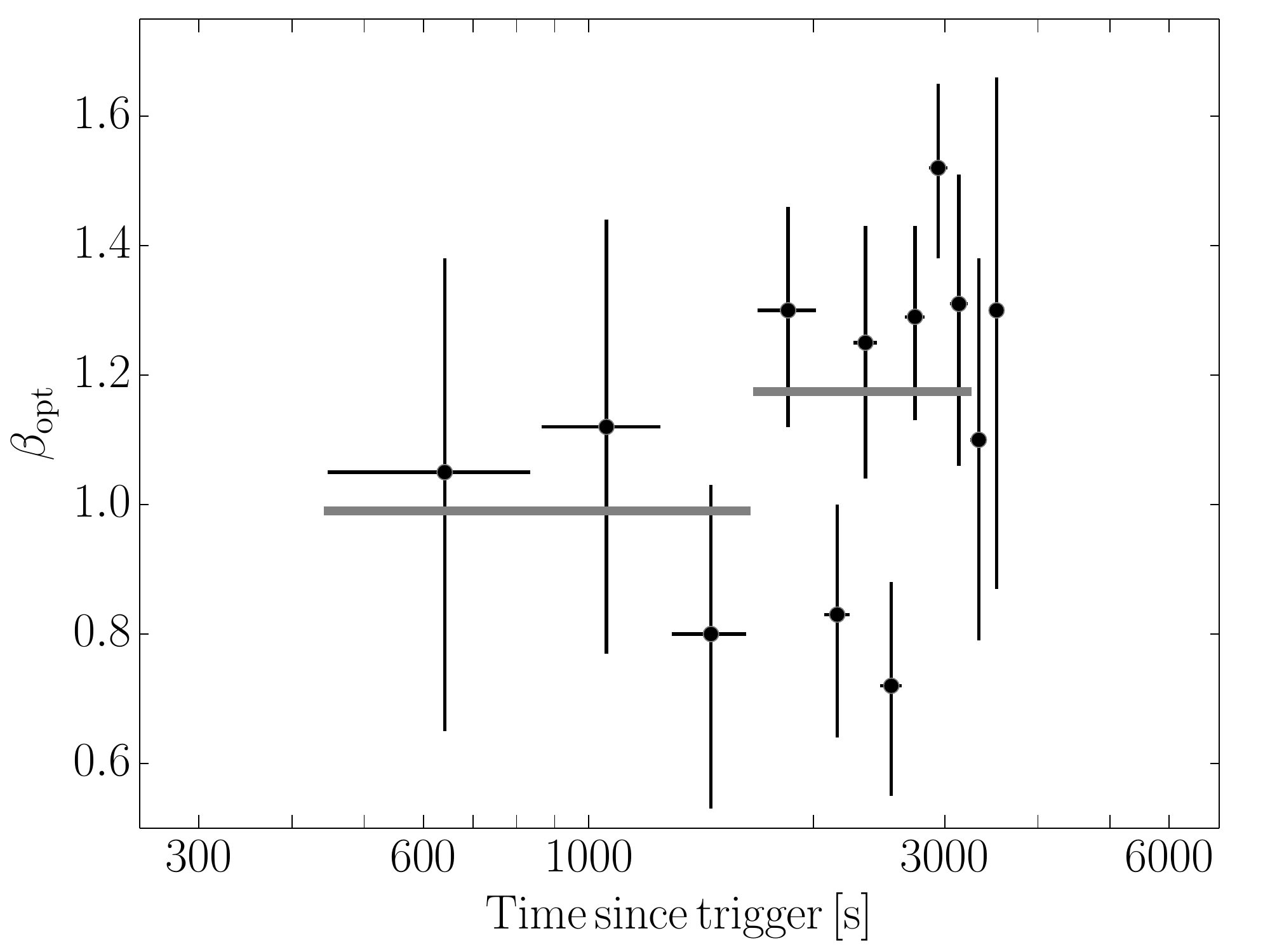}
\caption{{\bf Top:} Spectral energy distribution of GRB 130925A during the 
optical/NIR peak as observed with GROND. The emission of the host 
galaxy is subtracted.
There is no detection in the $g'r'$-bands above the host emission.
The curvature in the SED is due to the strong dust extinction
of $A_V = 5.0\pm0.7$ mag.
The insert shows contours of constant $\chi_{\rm red}$ for 
the intrinsic spectral slope as a function of reddening
(SMC reddening law $A_V = 2.93 \times E(B-V)$) for the peak.
{\bf Bottom:} Evolution of the spectral slope (spectral index 
$\beta$ which relates to the photon index via $\beta + 1$) of the 
optical/NIR afterglow with time with fixed $A_V$. 
For most temporal slices outside the optical/NIR peak, the
spectral energy distribution is defined only through $JHK_{\rm s}$ measurements,
and thus the error is relatively large. 
The horizontal bars visualize the average slope for the peak (right) and
the pre-peak (left) period.
\label{grondSED}
}
\end{figure}

When split into
single 10\,s integrations around the peak (only possible in the $K_s/H$-bands),
we find some substructure in the peak (Fig. \ref{delay}). 
To some extent, this substructure reflects the sub-pulse structure
in the KW light curve, in particular the first three sub-pulses 
(SP1--SP3 in Fig. \ref{delay}).
However, other features are missing or are less pronounced: the fourth
peak (SP4) is completely missing, the width of SP3 is much broader
in $K_s$ than in gamma-rays, while that of SP2 is narrower.
Also, the smaller peaks at 2000--2200 s after the BAT trigger are
hardly seen at all in $K_s$.
Moreover, a simple shift of the GROND $K_s$-band light curve (blue line
in Fig. \ref{delay}) is a better match than a time-stretch (green line).
We also note that the best-fit shift in Fig. \ref{delay} is by --290 s,
thus matching the peak of the $K_s$-band light curve with the sharp spike
SP3, while the above mentioned shift of 405$\pm$30 s matches the
highest fluence peak SP2.
We note the interesting property of the optical/NIR flare, namely that 
its decay time
of $\sim$20 s is much shorter than both the flare duration ($\sim$500 s) 
and its delay after the second peak ($\sim$300--400 s). The implications
of this are discussed further below.

\subsubsection{The spectral energy distribution}

The measured 5\,s peak magnitude in the $K$-band is 15.95 (AB) or 14.09 (Vega),
and after correction of the host-intrinsic extinction amounts to 
$K^{corr}$(Vega)=13.5 mag. This corresponds to a peak luminosity just in the
$K$ band of 3$\times$10$^{45}$ erg s$^{-1}$. The energy release of the main peak
(250\,s duration) is about 7$\times$10$^{48}$ erg in the 400--2400\,nm band.

The spectral energy distribution of the flare emission as measured with
GROND is very steep, but clearly curved, suggesting substantial extinction.
A powerlaw fit with SMC-like extinction results in an
extinction $A_{\rm V} = 5.0 \pm 0.7$ mag and a power law spectral
slope of $\beta_{opt}$ = 1.3 $\pm$ 0.4.
The relatively large errors stem from the correlation between these 
two parameters, see Fig. \ref{grondSED}. In any case, 
GRB 130925A is being extinguished by one of the highest dust columns
ever measured  \citep[e.g.,][]{gkk11}.

In an attempt to possibly distinguish canonical afterglow emission
from that of the optical/NIR peak, we have performed a time-resolved
spectral fit of the GROND data using the same time bins as for the
combined GROND-KW spectral fits (see below, and Tab. \ref{KWtime}).
With the extinction $A_{\rm V}$ 
fixed, the resulting best-fit spectral index $\beta$ is shown
in the bottom panel of Fig. \ref{grondSED}.
We find that the spectral slope
during the optical/NIR peak is steeper by at least 0.3 at the 2$\sigma$ level.

\subsubsection{Combined GROND and Konus-Wind spectral fitting}

Since the GROND data have less flexibility in rebinning than the Konus-Wind
data, we re-binned the KW data into time bins as determined by the
GROND exposures. We define 11 time slices as given in Tab. \ref{KWtime}.
Using the method described in section 2.1, 
we extracted a set of background-subtracted 3-channel spectra 
for these 11 time slices, and created a corresponding detector
response matrix using standard KW tools.

We then performed a combined fit of the  3-channel KW spectra and
the 3--5 (depending on significant excess emission above the host galaxy) 
filter GROND SED by employing a Band function \citep{bmf93}. Fixing the
redshift, and requiring the same extinction in all 11 fits,
the resulting best-fit parameters are also shown in  Tab. \ref{KWtime}.
While the fits are statistically acceptable 
(reduced $\chi^2$=34.1/34 for 79 bins and 34 degrees of freedom for
the combined fit of all 11 spectra together), 
the resulting parameters
are physically questionable. With a few exceptions, $E_{\rm peak}$
is always at the low-energy boundary of the G2 channel. 
Furthermore, the low-energy slope is always close to 1, with the variations
reflecting the relative rise and fall of the optical/NIR or $\gamma$-ray 
flux. Thus, while the model mathematically fits, we do not interpret
this as evidence that the optical/NIR emission tracks the $\gamma$-ray
emission.

\begin{table*}[th]
   \caption{Sequence of Konus-Wind spectra and results of joint GROND-KW 
   spectral fits:
  alpha and beta are the photon indices of the Band function, and 
  the normalization is at 100 keV. Start and stop times are seconds
  of the day.}
   \vspace{-0.2cm}
      \begin{tabular}{ccccccc}
      \hline
      \noalign{\smallskip}
      Spectrum  & start UT & stop UT & alpha & beta  & Epeak & Norm  \\
                &   (s)    &  (s)    &       &       & (keV) & (10$^{-3}$ ph/cm$^2$/s/keV) \\
      \noalign{\smallskip}
      \hline
      \noalign{\smallskip}
 1 & 16754.190 & 17113.358 & 1.08$\pm$0.01 & 2.38$\pm$0.25 & 106$\pm$25 & 5.5$\pm$0.4 \\
 2 & 17154.574 & 17319.438 & 1.00$\pm$0.01 & 2.64$\pm$0.22 & 114$\pm$11 & 14.6$\pm$0.6\\
 3 & 17351.822 & 17513.742 & 0.99$\pm$0.01 & 2.45$\pm$0.14 & 107$\pm$10 & 18.3$\pm$0.7\\
 4 & 17546.126 & 17710.990 & 0.94$\pm$0.01 & 2.44$\pm$0.07 & 117$\pm$5 & 35.7$\pm$0.8 \\
 5 & 17740.430 & 17902.350 & 1.01$\pm$0.01 & 2.66$\pm$0.17 & 110$\pm$8 & 20.9$\pm$0.7 \\
 6 & 17943.566 & 18105.486 & 1.22$\pm$0.03 & 4.03$\pm$52.5 & 56$\pm$61 & 2.4$\pm$0.9 \\
 7 & 18137.870 & 18299.790 & 1.15$\pm$0.02 & 9.38$\pm$14.8 & 72$\pm$35 & 2.8$\pm$0.6 \\
 8 & 18335.118 & 18499.982 & 1.03$\pm$0.05 & 4.72$\pm$33.7 & 48$\pm$52 & 7.4$\pm$3.8 \\
 9 & 18523.534 & 18688.398 & 0.98$\pm$0.02 & 3.72$\pm$3.31 & 92$\pm$19 & 9.4$\pm$0.9 \\
10 & 18729.614 & 19088.782 & 1.06$\pm$0.02 & 3.79$\pm$3.66 & 92$\pm$21 & 3.9$\pm$0.4 \\
11 & 19118.222 & 19477.390 & 1.00$\pm$0.02 & 2.89$\pm$0.68 & 91$\pm$18 & 6.1$\pm$0.5 \\
     \noalign{\smallskip}
      \hline
   \end{tabular}
   \label{KWtime}
\end{table*}

\begin{figure}[ht]
\includegraphics[width=9.5cm]{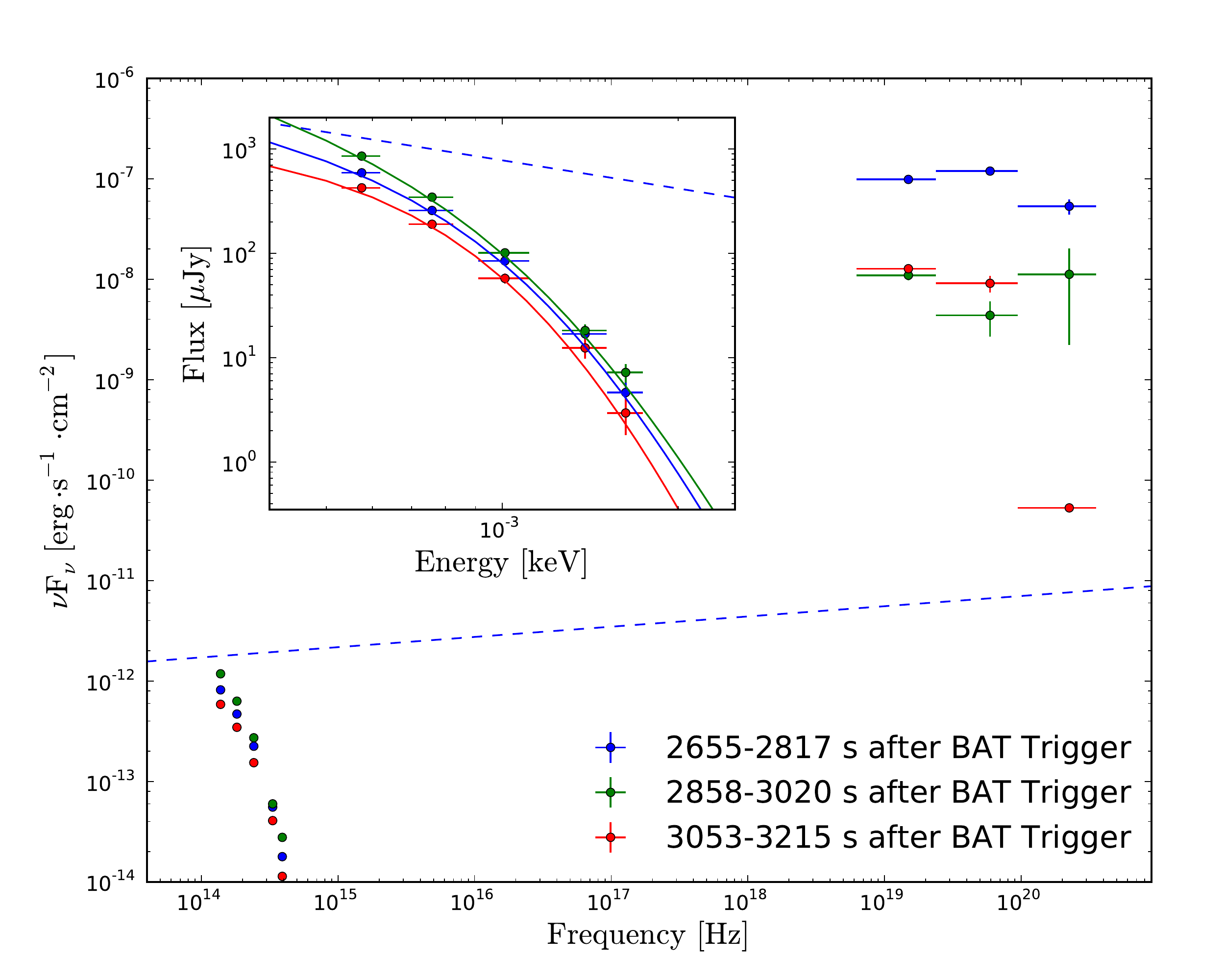}
\caption{Broad-band SED at the three epochs 5, 6, and 7
during the optical/NIR flare
as seen with GROND (left part) when the prompt flux as seen by 
Konus-Wind (right part) is tailing off into background emission. 
The inset shows a zoom-in to the GROND data and the best-fit 
power-law model (dashed: un-extinguished power law; solid: 
power law after extinction).
\label{GRONDKWspec}}
\end{figure}

\section{Discussion}

Figs. \ref{KWlc} and \ref{lc} suggest two possible
alternative interpretations. In the first, one would argue that the optical/NIR 
flare has occurred between the second and third gamma-ray emission episodes,
and thus is part of the prompt emission, similar to the two X-ray flares
before the optical/NIR flare. In this interpretation, one would need to
explain why the spectral peak of the various emission episodes changes 
so drastically between gamma-rays and optical. 
In a second interpretation, one could consider the optical/NIR flare as
delayed emission with respect to the second gamma-ray episode, in which case the
challenge is to explain both the delay as well as the fast decay time.
We discuss each of these options below.

\subsection{The optical/NIR flare as part of the prompt emission}

\subsubsection{Hard-to-soft evolution}

Analysis of energy-resolved gamma-ray light curves typically shows
a generic hard-to-soft evolution of peaks throughout a GRB,
manifesting in different features: 
(i) later peaks have smaller peak energies,
(ii) soft energies lag hard energies \citep{Cheng95, nmb2000},
(iii) the duration $T_{90}$ gets larger with energy according to
  $E^{-0.4}$ \citep{Biss11}.
In the case of GRB 130925A, the optical/NIR peak shows a delay which
is consistent with a $E^{-0.4}$ dependence, but the
widths of the peaks (or $T_{90}$) certainly do not.
We therefore conclude that the observed optical/NIR peak is likely 
not a manifestation of the hard-to-soft evolution of GRB pulses,
despite some similarity in the substructure of the emission.

\subsubsection{A canonical flare}

\begin{figure}[th]
\hspace{-0.3cm}\includegraphics[width=9.5cm]{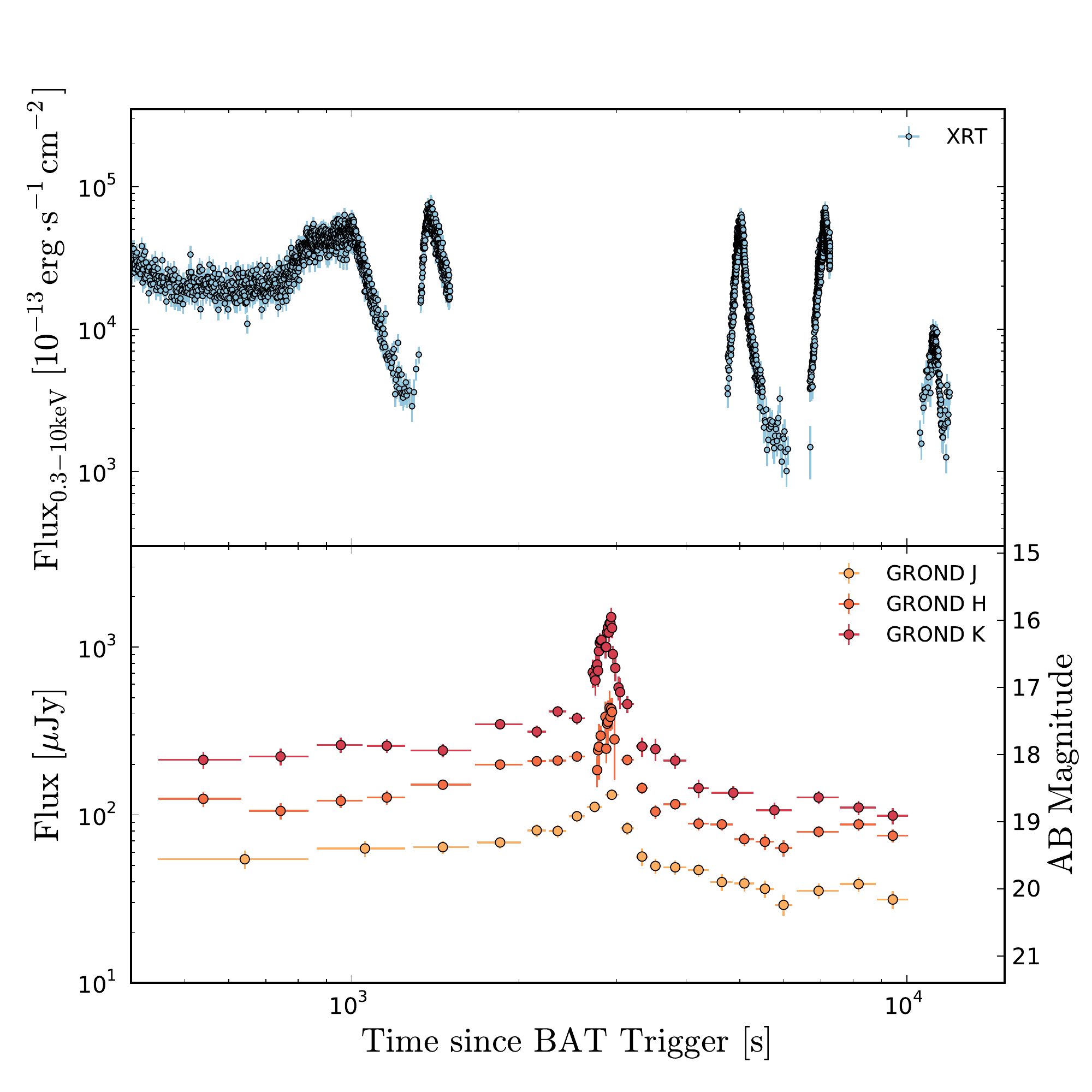}
\caption{
\textit{Swift}/XRT X-ray 
and GROND $JHK_{\rm s}$ host-subtracted light curve; while the GROND peak
is not covered by \textit{Swift}, the X-ray flares do not show up in the 
GROND data.
\label{lcwHST}
}
\end{figure}

Looking at Fig. \ref{lcwHST} shows that the observed optical/NIR flare
is temporally just inbetween major X-ray flares as seen by \textit{Swift}/XRT.
However, the optical/NIR flare seen with GROND is very unlikely to be
the optical ``counterpart'' of a missed X-ray flare, because the times
of three X-ray flares seen at 1,2\,ks, 5\,ks and 7\,ks are covered by 
GROND data, and show no sign of flux enhancement. Also, the rise and
decay times are different, namely $\sim$500\,s in X-rays as compared
to 20--30\,s in the optical/NIR.

\subsubsection{Reverse shock}

The reverse shock is predicted to happen with little delay with respect to
the gamma-ray emission unless the Lorentz factor is very small.
The peak of the emission occurs when the reverse shock crosses the
blast wave shell \citep{mer97, kob00}.
The corresponding optical emission has a rise time power law index of
+0.5 for both a wind and constant density profile (except for the
hypothetical case where self-absorption is relevant in the optical
regime, when the rise can be as fast as +2.5), and a
decay time power law index of 
about --2 for a constant density environment, or up to --3.0 for
a wind density profile \citep{kob00, koz03}.
The decay slopes are clearly inconsistent with our observations.
The rise slope could be consistent with the self-absorbed case,
but we observe a very sharp peak, so there is no flattening
towards a non-self-absorbed phase with a +0.5 rise.
We therefore exclude a reverse shock interpretation of the optical/NIR
flare in GRB 130925A.

\subsection{The optical/NIR flare as delayed emission}

\subsubsection{Dust destruction}

A number of suggestions have been made in the past which relate
the delayed onset of optical/UV emission to the effect of dust
in the nearby surrounding medium. 

One scenario invokes a dense region around the GRB progenitor,
similar to that of a molecular cloud. While the prompt gamma-rays
and the X-ray afterglow would pass relatively unaffected, the
optical emission would only pass once the dust along our line of sight
is completely destroyed through thermal sublimation \cite[e.g.][]{Cui2013}.
In the case of GRB 130925A, this scenario is ruled out, because
we observe no reduction of the best-fit extinction throughout 
the burst, to a limit of $\Delta A_{\rm V} < 0.4$ mag. 

\subsubsection{Pair-loaded fireball}

Another more promising possibility seems to be the framework of the e${^\pm}$ 
enrichment originally proposed to explain the delayed GeV flash 
in \textit{Fermi}/LAT data \citep{bhv13}. The prompt MeV radiation
streaming ahead of the blast wave creates a large enrichment in e$^\pm$ 
which then in turn leads to substantial inverse Compton scattering
of the prompt MeV photons, thus creating a GeV flash. This GeV emission
starts while the prompt emission is still going on, but lasts longer
than the prompt emission because of a broader angular distribution 
of scattered photons. The model reproduces the delayed onset, 
the steep rise, the peak flux, the time of the peak, the long smooth 
decay and the spectral slope of the GeV flash emission \citep{bhv13}.
Simultaneously, a bright optical flash is 
predicted \citep{bhv13}, which has been nicely confirmed 
in the RAPTOR data \citep{vwp14} of the GRB 130427A afterglow \citep{vhb14}.

The delay of the GeV/optical flash is determined by two parameters,
the pair-loading radius and the Lorentz factor \citep{bhv13}. 
The pair-loading radius
is $R_{\rm load} \approx 10^{17} E^{1/2}_{\rm 54}$ cm, where
$E_{\rm 54}$ is the isotropic equivalent energy of the prompt GRB energy ahead
of the forward shock. For GRB 130925A, we take 
$E_{\rm 54}$ as the sum of the first prompt pulse and half of the
second, i.e. $E_{\rm 54}=0.08$, and thus 
$R_{\rm load} = 2.9\times 10^{16}$ cm. Together with the 
delay of 3040 / 405 s (relative to first or second episode) 
we obtain an inferred Lorentz factor of
$\Gamma_{405s} \approx ( (1+z)*R_{\rm load}/2c\delta t)^{1/2} \sim 40$
or $\Gamma_{3040s} = 11$.
It is not obvious which $\gamma$-ray peak to take as reference:
One could argue that the delay of the observed optical/NIR peak
should be taken relative to the first emission episode, as the
blast wave should start immediately after the beginning of the
explosion (possibly even at the precursor). However, each prompt
emission peak is likely associated with a large energy injection 
into the blast wave, which may effectively offset the reference
time. 
In any case, the inferred Lorentz factors are of the same order 
of magnitude as the Lorentz factor 
derived above from the rising part of the afterglow,
and the low absolute value of the inferred Lorentz factor
is also nicely compatible with the exceptionally long minimum 
variability time scale.

As a consequence of the low Lorentz factor the corresponding
photosphere of the GRB ejecta \citep{hbd14}
implies a photospheric time scale of
1.9 ms ($\Gamma=40$) and 1.2 s ($\Gamma=11$) for the above two
options. The latter Lorentz factor is perfectly consistent with the MVT 
of one and
five seconds as deduced from the prompt emission of the first and
second emission episodes.

The only potential mismatch with this pair-loaded fireball scenario 
is the sub-structure in
the flare, as well as its fast decay time scale. It remains to be seen
through detailed simulations whether this scenario will work 
for GRB 130925A.

\subsubsection{Internal dissipation}

Finally, the fast decay of the optical/NIR flare could be considered
as evidence for an internal origin within the jet, similar to models for 
the X-ray flares (but see above section 4.1.). 
In the standard picture of the blast-wave emission the curvature
effect implies that the decay time scale of a flare is
of the same order of magnitude as its duration. This can be avoided
either by a jet opening angle smaller than 1/$\Gamma$, or 
by non-isotropic emission in the blast-wave frame 
\citep[e.g., limb-brightened][]{bdm11}. 
The former case is unlikely, as this GRB shows neither a
particularly spiky light curve, nor evidence of particularly high
Lorentz factor. In contrast, 
the latter case in fact also produces a delay of
the emission at very small spreading in time, exactly as we observe 
in GRB 130925A.
If this is the true interpretation, GRB 130925A might be the first
observational evidence for limb-brightened jet emission.

If the optical/NIR flare and the second gamma-ray peak are indeed
related, e.g. by being different episodes of emission from the same
shell but at different radii, one would expect the GROND light curve to
be stretched rather than only shifted with respect to the second gamma-ray 
peak (as Fig. \ref{delay} demonstrates, both options are consistent with 
the data, with a preference for a shift). 
The similarities between the two peaks are then likely the result
of the imprint of the outflow geometry or jet structure on the emission,
which might be dominated by different processes at the different
frequencies.

\subsection{Comparison to previous GRBs}

While there have been very early optical observations covering the
prompt gamma-ray emission for about a dozen objects, the combination of 
sensitivity and time-resolution has in most cases not allowed the 
establishment of a clear timing pattern of the optical relative to the
gamma-ray emission. Apart from GRB 130427A, which is explained in 
a different way \citep{vhb14}, the most obvious cases showing
substantial delay of the optical flare relative to the gamma-ray
emission are 
GRB 041219A as observed with PAIRITEL \citep{blake05},
one of the other few ultra-long bursts GRB 111209A as observed by 
TAROT \citep{kgb11, sga13},
and GRB 121217A as observed with GROND \citep{Elliott14}.
In all these cases, the optical flare occurs delayed, but still within the
T90 of the prompt gamma-ray emission.
Less clear cases wrt delay, though also with clear optical variability,
are GRBs 080129 \citep{gkm09} and 080928 \citep{rossi11}. 
The interpretation of these lags has been diverse: while \cite{blake05}
argue that the $JHK_{\rm s}$ emission is consistent with internal shocks,
\cite{sga13} suggest two different emission regions, and
\cite{rossi11} discuss large-angle emission where the synchrotron
peak flux and the peak energy of the electron ensemble are functions of
the viewing angle of the observer with respect to the jet axis.
GRB  041219 is
the only earlier flare which also showed sub-structure, and thus has
a similar interpretative problem as our GRB 130925A.

The very long minimum variability time, the relatively faint afterglow
with respect to the prompt emission, and the very shallow decay of
the afterglow up to very late times suggest a
low Lorentz factor, of order 10--20.
This is interesting with 
respect to the range of Lorentz factors of 30--400 recently deduced for
a sample of GRBs \citep{hbd14}. This suggests that the range of
Lorentz factors could be even broader, 
with little effect on the prompt GRB spectrum
and/or luminosity.
We note in passing that this would also be consistent with the photospheric
origin of the prompt emission in GRBs 
\citep{thom94, RyPe09, Peer11, rpn11, Gian12, vlp13, rsv13, bel13}.

\section{Conclusions}

We have observed a pronounced  optical/NIR flare related to GRB 130925A.
Our well-sampled multi-color GROND light curve shows a peak
which is delayed by 290 s with respect to the second $\gamma$-ray peak. 
This is not dissimilar to several other GRBs for
which prompt optical emission has been detected. However, the exquisite
data quality for GRB 130925A allows us to exclude the leading contenders.
First, given the delay by nearly the full pulse-width, the optical/NIR
emission cannot be just the extrapolation of a Band (or any other
broad-band model) extension from the $\gamma$-ray emission.
Second, the typical interpretation of being due to a reverse shock also
does not apply, since the  
optical/NIR peak in  GRB 130925A exhibits a very sharp rise and an 
extremely steep decay, both much faster than expected for a reverse shock.
Other options have been discussed as well, suggesting that the main
observational feature of the decay time being shorter than the delay time
can, in principle, be explained. However, much more detailed considerations
are needed to derive a coherent picture, which are beyond the scope of
this paper.

\begin{acknowledgement}
We thank the referee, D. Burrows, for a careful reading of the
manuscript and his detailed suggestions.
JG and HFY acknowledge support by the DFG cluster of excellence 
``Origin and Structure of the Universe'' (www.universe-cluster.de).
PS, JFG and MT acknowledge support through the Sofja Kovalevskaja award
to P. Schady from the Alexander von Humboldt Foundation Germany. 
SK and ANG acknowledge support by DFG grant Kl 766/16-1.
SS acknowledges support by the Th\"uringer Ministerium f\"ur Bildung,
Wissenschaft und Kultur, FKZ 12010-514.
DAK acknowledges TLS Tautenburg for financial support.
AvK and XLZ acknowledge support by Space Agency of the Deutsches
Zentrum f\"ur Luft- und Raumfahrt e.V. (DLR) through funding by
the Bundesministerium f\"ur Wirtschaft und Technologie under FKZ 
50 OG 1101.
KH is grateful for support under NASA grants NNX13AI54G, NNX11AP96G, and
NNX07AR71G.
Part of the funding for GROND (both hardware as well as personnel)
was generously granted from the Leibniz-Prize to Prof. G. Hasinger
(DFG grant HA 1850/28-1).
The Konus-$WIND$ experiment is partially supported by a Russian 
Space Agency contract, RFBR grants 12-02-00032a and 
13-02-12017 ofi-m.
This work made use of data supplied by the UK \textit{Swift} Science Data
Centre at the University of Leicester.

\end{acknowledgement}

\bigskip


\noindent {\small {\it Facilities:} Max Planck:2.2m (GROND),  
                  Konus-Wind, INTEGRAL (SPI-ACS), Swift, Suzaku}


\begin{thebibliography}{}

\bibitem[Akerlof et al.(1999)]{akerlof99} Akerlof C., Balsano R., 
Barthelmy S.,  \etal\ 1999,  Nature 398, 400

\bibitem[Aptekar et al.(1995)]{Aptekar1995} Aptekar, R.L., Frederiks, D.D., 
Golenetskii, S.V., et al.\ 1995, SSR 71, 265

\bibitem[Band et al.(1993)]{bmf93} Band D., Matteson J., Ford L. et al. 
1993, ApJ 413, 281

\bibitem[Blake et al.(2005)]{blake05} Blake C.H., Bloom J.S., Starr D.L. 
et al. 2005, Nat. 435, 181

\bibitem[Beloborodov et al.(2011)]{bdm11} Beloborodov A.M., Daigne F.,
  Mochkovitch R., Uhm Z.L., 2011, MN 410, 2422

\bibitem[Beloborodov(2013)]{bel13} Beloborodov A.M., 2013, ApJ 764, A157

\bibitem[Beloborodov et al.(2014)]{bhv13} Beloborodov A.M.,
  Haso\"et R., Vurm I., 2014, ApJ 788, A36 

\bibitem[Bhat(2013)]{Bhat13} Bhat P.N., 2013, in ``Proc. 7th Huntsville 
GRB Symp.'', Huntsville, Apr. 2013, eConf Proc. C1304143

\bibitem[Bissaldi et al.(2011)]{Biss11} Bissaldi E., von Kienlin A., 
Kouveliotou C. et al. 2011, ApJ 733, 97

\bibitem[Bj\"ornsson et al.(2004)]{bgj04} Bj\"ornsson G., Gudmundsson E.H., 
  Johannesson G., 2004, ApJ 615, L77

\bibitem[Blake et al.(2005)]{bbs05} Blake C.H., Bloom J.S., Starr D.L., et al. 
  2005, Nat. 435, 181

\bibitem[Burrows et al.(2011)]{Burrows11} Burrows D.N., Kennea J.A.,
 Ghisellini G., et al. 2011, Nat. 476, 421

\bibitem[Burrows et al.(2013)]{Burrows13} Burrows D.N., Malesani D.,
Lien A.Y., Cenko S.B., Gehrels N., 2013, GCN 15253

\bibitem[Cheng et al.(1995)]{Cheng95} Cheng L.X., Ma Y.Q., Cheng K.S.,
 Zhou Y.Y., 1995, A\&A 300, 746

\bibitem[Cui et al.(2013)]{Cui2013}
Cui X.-H., Li Z., Xin L.-P., 2013, Res. in Astron. Astrophys. 13, No. 1, 57

\bibitem[de Ugarte Postigo et al.(2005)]{ucg05} de Ugarte Postigo A.,
Castro-Tirado A.J., Gorosabel J. et al. 2005, A\&A 443, 841

\bibitem[Elliott et al.(2014)]{Elliott14} Elliott J., Yu H.-F., 
Schmidl S., et al. 2014, A\&A 562, A100

\bibitem[Evans et al.(2009)]{ebp09} Evans, P.A., Beardmore, A.P., Page, K.L.,
   et al. 2009, MN 397, 1177

\bibitem[Evans et al.(2014)]{Evans14}  Evans P.A., Willingale R., Osborne J.P.
et al. 2014, MN (subm., arXiv:1403.4079)

\bibitem[Fan et al.(2009)]{Fan2009} Fan Y., Zhang B., Wei D., 2009, 
PhRv 79, 021301

\bibitem[Fitzpatrick(2013)]{Fitzpatrick13} Fitzpatrick G., 2013, GCN \#15255

\bibitem[Gehrels et al.(2004)]{gcg04} Gehrels N., Chincarini G., Giommi P., 
et al. 2004,    ApJ 621, 558


\bibitem[Ghirlanda et al.(2012)]{gng12} Ghirlanda G., Nava L., Ghisellini G.
  et al. 2012, MN 420, 483

\bibitem[Giannios(2012)]{Gian12} Giannios D., 2012, MN 422, 3092

\bibitem[Golenetskii et al.(2013)]{Golenetskii13} Golenetskii S., 
Aptekar R.,  Frederiks D. et al. 2013, GCN \#15260

\bibitem[Greiner et al.(2008)]{gbc08}
Greiner J., Bornemann W., Clemens C., et al. 2008, PASP 120, 405

\bibitem[Greiner et al.(2009)]{gkf09} Greiner J., Kr\"uhler T., 
Fynbo J.P.U, et al. 2009, ApJ 693, 1610

\bibitem[Greiner et al.(2009)]{gkm09} Greiner J., Kr\"uhler T., McBreen S., 
et al. 2009, ApJ 693, 1912

\bibitem[Greiner et al.(2011)]{gkk11}
 Greiner J.,  Kr\"uhler T., Klose S. et al. 2011, A\&A 526, A30

\bibitem[Hasco\"et et al.(2014)]{hbd14} Hasco\"et R., Beloborodov A.M., 
Daigne F., Mochkovitch R., 2014, ApJ 782, A5

\bibitem[Hurley et al.(2013)]{hga13} Hurley K., Golenetskii S., Aptekar R.,
et al. 2013, GCN \#15278

\bibitem[Kalberla(2005)]{kalberla2005} Kalberla, P.M.W., Burton, W.B., 
Hartmann, D., et al.\ 2005, A\&A 440, 775

\bibitem[Klotz et al.(2006)]{kgs06} Klotz A., Gendre B., Stratta G., et al. 
 2006, A\&A 451, L39

\bibitem[Klotz et al.(2011)]{kgb11} Klotz A., Gendre B., Boer M., 
  Atteia J.-L., 2013, GCN \#12637

\bibitem[Kobayashi(2000)]{kob00} Kobayashi S., 2000, ApJ 545, 807

\bibitem[Kobayashi \& Zhang(2003)]{koz03} Kobayashi S., 
Zhang B., 2003, ApJ 597, 455

\bibitem[Kr\"uhler et al.(2008)]{kkg08} Kr\"uhler T., K\"{u}pc\"{u} Yolda\c{s}
 A., Greiner J.,   et al. 2008, ApJ 685, 376

\bibitem[Kr\"uhler et al.(2011)]{ksg11} Kr\"uhler T., Schady P., Greiner J., 
et al. 2011, A\&A 526, A153

\bibitem[K\"{u}pc\"{u} Yolda\c{s} et al.(2008b)]{kkg08b} K\"{u}pc\"{u} Yolda\c{s} A., Kr\"uhler T., Greiner J., 
 et al. 2008b, AIP Conf. Proc., 1000, 227

\bibitem[Lazzati et al.(2002)]{lrc02} Lazzati D., Rossi E., Covino S., 
Ghisellini G., Malesani D. 2002, A\&A 396, L5

\bibitem[Levan et al.(2011)]{Levan11} Levan A.J., Tanvir N.R., 
Cenko S.B., et al. 2011, Sci 333, 199

\bibitem[Li \& Waxman(2008)]{liw08} Li Z., Waxman E., 2008, ApJ 674, L65

\bibitem[Liang et al.(2010)]{lyz10} Liang E.-W., Yi S.-Y., Zhang J. 
et al. 2010, ApJ 725, 2209

\bibitem[Lien et al.(2013)]{Lien13} Lien A.Y., Markwardt C.B., 
Page K.L. et al. 2013, GCN \#15246

\bibitem[Lipkin et al.(2004)]{log04} Lipkin Y.M., Ofek E.O., Gal-Yam A.,
et al. 2004, ApJ 606, 381

\bibitem[Lithwick \& Sari(2001)]{lis01} Lithwick Y. \& Sari R., 2001, ApJ 
555, 540

\bibitem[Malesani et al.(2013)]{Malesani13} Malesani D., Cenko S.B., 
  Burrows D.N., Evans P.A., Gehrels N., 2013, GCN \#15280

\bibitem[Maselli et al.(2014)]{Maselli14} Maselli A., Melandri A., Nava L. 
et al. 2014, Science 343, 48

\bibitem[Melandri et al.(2010)]{mkm10} Melandri A., Kobayashi S., 
Mundell C.G. et al. 2010, ApJ 723, 1331

\bibitem[Meszaros \& Rees(1997)]{mer97} Meszaros P., Rees M.J., 
1997, ApJ 476, 232

\bibitem[Meszaros \& Rees(1999)]{mer99} Meszaros P., Rees M., 1999, MN 306, L39

\bibitem[Meegan et al.(2008)]{mlb08} Meegan C., Lichti G., Bhat P.N. 
et al. 2008, ApJ 702, 791 

\bibitem[Norris et al.(2000)]{nmb2000} Norris J.P., Marani G.F., Bonnell J.T.,
2000, ApJ 534, 248

\bibitem[Oates et al.(2009)]{Oates09} Oates S.R., Page M.J., Schady P. et al.
  2009, MN 395, 490

\bibitem[Pe'er(2011)]{Peer11} Pe'er A., 2011, ApJ 732, A49

\bibitem[Piro et al.(2014)]{Piro14} Piro L., Troja E., Gendre B., et al. 2014,
ApJ (subm., arXiv:1405.2897)

\bibitem[Racusin et al.(2008)]{rks08} Racusin J.L., Karpov S.V., Sokolowski M.,
  et al. 2008. Nat. 455, 183

\bibitem[Rossi et al.(2011)]{rossi11} Rossi A., Schulze S., Klose S. et al.
 2011, A\&A 529, A142

\bibitem[Ruffini et al.(2013)]{rsv13} Ruffini R., Siutsou I.A., 
Vereshchagin G.V., 2013, ApJ 772, A11

\bibitem[Ryde \& Pe'er(2009)]{RyPe09} Ryde F., Pe'er A., 2009, ApJ 702, 1211

\bibitem[Ryde et al.(2011)]{rpn11} Ryde F., Pe'er A., Nymark T., et al.
 2011, MN 415, 3693

\bibitem[Rykoff et al.(2004)]{ryk04} Rykoff E., et al. 2004, ApJ 601, 1013

\bibitem[Sari \& Piran(1999)]{sap99} Sari R., Piran T., 1999, ApJ 517, L109

\bibitem[Savchenko et al.(2013)]{Savchenko13} Savchenko V., Beckmann V.,
Ferrigno C., et al. 2013, GCN \#15259

\bibitem[Schlegel et al.(1998)]{sfd98} Schlegel D., Finkbeiner D., Davis
 M. 1998, ApJ 500, 525

\bibitem[Stratta et al.(2013)]{sga13} Stratta G., Gendre B., Atteia J.L.
et al. 2013, ApJ 779, 66

\bibitem[Sudilovsky et al.(2013a)]{Sudilovsky13a} Sudilovsky V., 
 Kann D.A., Greiner J., 2013a, GCN \#15247 

\bibitem[Sudilovsky et al.(2013b)]{Sudilovsky13b} Sudilovsky V., 
 Kann D.A., Schady P., Klose S., Greiner J., Kr\"uhler T., 2013b, 
GCN \#15250 

\bibitem[Suzuki et al.(2013)]{Suzuki13} Suzuki K., Sakakibara H., 
Negoro H., et al. 2013, GCN \#15248

\bibitem[Swenson et al.(2013)]{Swenson13} Swenson C.A., Roming P.W.A., 
De Pasquale M., Oates S.R., 2013, ApJ 774, A2

\bibitem[Tanvir et al.(2013)]{Tanvir13} Tanvir N.R., Levan A.J., 
Hounsell R., et al. 2013, GCN \#15489 

\bibitem[Thompson(1994)]{thom94} Thompson C., 1994, MN 270, 480

\bibitem[Tody(1993)]{Tody1993} Tody D., 1993, in ASP Conf. 52,
Astronomical Data Analysis Software and Systems II, ed. R.J. Hanisch,
R.J.V. Brissenden, \& J. Barnes, p. 173

\bibitem[Vestrand et al.(2005)]{vest05} Vestrand W.T., et al. 2005, 
 Nat. 435, 178

\bibitem[Vestrand et al.(2006)]{vest06} Vestrand W.T., et al. 2006, Nat. 442, 172

\bibitem[Vestrand et al.(2014)]{vwp14} Vestrand W.T., Wren J., Panaitescu A.
  et al. 2014, Sci. 343, 38

\bibitem[Vreeswijk et al.(2013)]{Vreeswijk13} Vreeswijk P.M.,
Malesani D., Fynbo J.P.U., De Cia A., Ledoux C., 2013, GCN \#15249

\bibitem[Vurm et al.(2013)]{vlp13} Vurm I., Lyubarsky Y., Piran T., 2013,
ApJ 764, A143

\bibitem[Vurm et al.(2014)]{vhb14} Vurm I., Hasco\"et R., Beloborodov A.M.,
2014, ApJ (subm.; arXiv:1402.2595)

\bibitem[Zhang et al.(2014)]{Zhang14} Zhang B.-B., Zhang B., Murase K.,
Connaughton V., Briggs M., 2014, ApJ 787, 66

\bibitem[Zhao \& Shao(2014)]{Zhao14} Zhao Y.-N., Shao L., 2014, ApJ (subm.,
arXiv:1403.3825)

\end{thebibliography}
\end{document}